\documentclass[smallabstract,smallcaptions]{dccpaper}
\usepackage{amsmath,amssymb,graphicx,booktabs,url,placeins}
\newcommand{\Wl}{W(\ell)}
\begin{document}

\title{\large\textbf{WZIP and WLZ4: LZ77 Codecs with\\Match-Length-Dependent Sliding Windows}}

\author{%
Yingquan (Cody) Wu\\[0.5em]
{\small\begin{minipage}{0.9\linewidth}\begin{center}
Tenafe Inc.\\
591 W. Hamilton Ave\\
Campbell, CA 95008, USA\\
\url{icodywu@gmail.com}
\end{center}\end{minipage}}
}

\maketitle
\thispagestyle{empty}

\begin{abstract}
An LZ77 codec bounds every match by one sliding window, trading cheap offsets for short matches against reach for
long ones. Match-length-dependent windows refine this trade-off: each match length has its own
window, and the decoder infers the offset range from the decoded length. We present two such codecs: WLZ4, a
byte-aligned LZ4-class format with one-byte and flagged offsets, and WZIP, an entropy-coded
Zstandard-class format. Each block begins with its decoded length, from which WZIP derives its widest
window and a decoder sizes its output. LZ4 blocks may expand incompressible input by 0.4\%; WLZ4 adds at most 15 bytes,
WZIP 2. We describe the formats, threaded match finders, decoders and parser. On Silesia, WLZ4 reaches a
ratio up to 16\% above LZ4HC's, decoding more slowly, and WZIP exceeds Zstandard 22's ratio while its trusted decoder runs at 91\%
of Zstandard's speed (87\% checked; on enwik9, about as fast).
\end{abstract}

\section{Introduction}
\label{sec:intro}

LZ77~\cite{LZ77} replaces a repeated substring by a pair $(\ell,d)$: copy $\ell$ bytes from $d$ bytes back. Every
practical LZ77 codec bounds $d$ by a window $W$, and the choice of $W$ is a compromise: a wide window exposes distant
repetitions but makes every offset expensive, so that a 3-byte match within a 64\,KiB window saves little or nothing,
while a narrow window misses long repetitions further back. Byte-aligned codecs such as LZ4~\cite{LZ4} fix the window
at 64\,KiB and the minimum match at 4 bytes; entropy-coded codecs such as DEFLATE~\cite{DEFLATE},
Zstandard~\cite{Zstd}, Brotli~\cite{Brotli} and LZMA/xz~\cite{LZMA,XZ} ease the compromise with variable-length
offset codes, and LZMA and Brotli condition them on the match length (2, 3, 4, 5+), which changes the offsets'
probabilities but not their range: every length still reaches the whole window.

The top of each class is hard to move. Zstandard at level 22 combines a 128\,MiB window, a binary-tree optimal parser
and FSE entropy coding~\cite{ANS} (its \texttt{--max} mode, new in version 1.5.7, compresses
enwik9 5.6\% more at 12 times the time and 6 times the memory~\cite{Zstd157}); the codecs we
measured that compress Silesia further, Brotli at level 11 and xz at $-9e$, rely on context-modeled literal coding and
decode 2.6 and 8.8 times slower. LZ4HC at level 12 is similarly the strongest byte-aligned parser at LZ4's decoding speed.
We aim to improve on both, keeping each class's decoding structure: WZIP, with Huffman coding only, exceeds the
ratio of Zstandard~22 by 0.8--1.1\% on Silesia at levels 11--12 while decoding at 90--91\% of its speed in trusted mode (86--87\% checked), and at
level 13 by 1.8\% on enwik9, also above xz, decoding about as fast; WLZ4 reaches ratios up to 16\% above LZ4HC~12's on Silesia and 29\% on enwik9, decoding
at most 32\% slower.

Match-length-dependent windows assign each length $\ell$ a window $\Wl$, nondecreasing in $\ell$. Short matches are
admitted only at short distances, where their offsets are cheap, and long matches reach the whole history; since
lengths are decoded before offsets, the decoder knows $\Wl$. Byte-aligned formats have tied offset fields to lengths
before: Snappy~\cite{Snappy} limits its one-byte-offset copies (11-bit offsets) to lengths 4--11, and
Lizard~\cite{Lizard} gives 16-bit offsets to matches of 4 bytes or more and 24-bit offsets only to matches of 16 or
more. WLZ4 instead gives every match longer than 3 bytes a flag that buys one more offset byte, so that the 8\,MiB
window opens at length 6 rather than being reserved to long matches. Our earlier abstract~\cite{WuDCC2021} studied a
nondecreasing schedule over all lengths with an LZ4-based prototype, and~\cite{WuIT} analyzes it: exact greedy parsing
under length-dependent windows remains universal when the full window starts at a length $o(\log W)$, and a
calibrated schedule never raises the minimum token cost over a single window. This paper is about engineering: two complete codecs, the design decisions the scheme
enables, and measurements against current codecs. New relative to~\cite{WuDCC2021} are the flagged offset, WZIP's
size-derived schedule and joint symbols, the parser, threads per index, and the evaluation.

\section{Length-dependent windows}
\label{sec:windows}

A match $(\ell,d)$ at input position $i$ copies $x_{i-d},\ldots,x_{i-d+\ell-1}$ to $x_i,\ldots,x_{i+\ell-1}$, possibly
overlapping itself. Under a schedule $\Wl$, $(\ell,d)$ is admissible if $d\le\Wl$; a constant $W$ is the single-window
case. The schedule binds matches with a new offset: a repeat offset, coded by its cache slot,
and WZIP's run token are not restricted. In a byte-aligned format the schedule appears as offset fields of different
sizes, selected in WLZ4 by the length code and a flag, in an entropy-coded format as offset alphabets that grow with
the length; either way the decoder needs no extra memory.
Both codecs start at length 3: a length-2 match costs WLZ4 as much as the two literals it replaces, and within 16
bytes it lowered WZIP\_S's ratio on 4--8\,KiB blocks by 0.01--0.08\%.

\emph{The input length in the stream.} A schedule may also depend on the input length $n$: a window wider than the
input is wasted, and one that covers it lets the longest matches reach everything. Both formats therefore begin with
$n$ (two bytes below 32\,KiB, four otherwise; a stored WZIP input begins with a zero instead). The size costs at most four
bytes and serves three purposes: WZIP derives a size-adapted base schedule $W_n(\ell)$ from it; the decoder allocates
its output before decoding; and decoding can be checked against a known end. Snappy also stores the size up
front~\cite{Snappy}; the LZ4 block format omits it, and LZ4 and Zstandard frames record it optionally.

The schedule also shapes the search~\cite{WuDCC2021}: each class of lengths gets its own index, keyed by a hash of its
minimum length and covering only the class's window. Indexes on short seeds stay short even in repetitive
data, long matches come from an index on long seeds over the whole window, and each
index searched nearest first yields the nearest offset for each length, the cheapest when offset fields have fixed
widths.

\section{WLZ4: a byte-aligned multi-window format}
\label{sec:wlz4}

\emph{Format.} A block starts with its decoded size (2 bytes below 32\,KiB, else 4), so that, unlike an LZ4 block, it
needs no size from the application. It holds sequences, each a token, a literal run and a
match, as in LZ4. The token's high nibble is the literal-run length; its low nibble is a match
code $c$ for length $3+c$ (Table~\ref{tab:wlz4codes}), code 15 for 18 plus an extension. Code 0 takes a one-byte
offset; every other code takes a \emph{flagged offset}: a little-endian field of $1+[c>2]$ bytes, or one more if its
low bit is set, the offset filling the remaining bits. A nibble value of 15 announces an
extension $e$: one byte if $e<252$, else $251+m$ followed by $e-252$ in $m$ bytes. A zero offset ends the block.

\begin{table}[t]
\begin{center}
\caption{\label{tab:wlz4codes}WLZ4 match codes (low nibble of the token) and offset fields. Offset 0 ends a block, so a
window of $2^k$ bytes reaches offsets up to $2^k-1$.}
{\small
\begin{tabular}{cccc}
\toprule
Code & Length & Offset field & Window \\
\midrule
0 & 3 & 1 byte & 256\,B \\
1, 2 & 4, 5 & flagged: 1 byte (flag 0) or 2 bytes (flag 1) & 128\,B or 32\,KiB \\
3--14, 15 & 6--17, $18+$ext. & flagged: 2 or 3 bytes & 32\,KiB or 8\,MiB \\
\bottomrule
\end{tabular}}
\end{center}
\end{table}

The flag is what makes the far window pay: an earlier variant spent match code 15 on far matches, which then always
needed a length extension, and compressed Silesia 2.6--2.8\% worse (Section~\ref{sec:ablation}). The extension code
keeps the size of an extension logarithmic, and every encoder writes a sequence only if the output stays within four
bytes of the input it covers, so incompressible input expands by at most 15 bytes, including the header, where an LZ4
block may add $n/255$, or 0.4\%.

\emph{Decoding.} The decoder reads four bytes $v$ at the offset position; with $f_c=[c>0]$ and $b_c=[c>2]$ from two
16-entry tables, the offset field has $1+b_c+(v\wedge f_c)$ bytes and the offset is $v\gg f_c$, masked: one rule for
every code, without a branch. Because the flag makes the position of the next
token depend on a byte just read, the decoder takes the next token from $v$ unless an extension follows,
keeping a second load off each sequence's critical path.

\emph{Encoders.} The fast and lazy levels probe two hash tables (3- and 5-byte hashes) and take a candidate of at
least 3 bytes within 256\,B, 5 bytes within 32\,KiB, or 6 bytes within 8\,MiB. The hash-chain levels 0--7 walk a 5-byte
chain over 64\,KiB, compare candidates by net gain (length less the offset bytes beyond two), and probe two
far positions: the uncapped previous position of the same 5-byte hash, and the previous position of the same 8 bytes
from a table sized to the input (at most $2^{20}$ entries; it alone moved level 6 on Silesia from 2.70 to 2.81). Levels 8--12 run the optimal parser (Section~\ref{sec:opt}) over 3-, 4- and
6-byte chains (256\,B, 64\,KiB, 8\,MiB). Each level reaches at most its own far window: 8\,MiB at levels 7 and 12, half
per level below.

\section{WZIP: an entropy-coded multi-window format}
\label{sec:wzip}

\emph{Format.} WZIP codes an input of 32\,KiB or more as one block; smaller inputs (7 of the 29 small files) go to
WZIP\_M, a variant with fixed 8 and 32\,KiB windows and two interleaved sequence streams, and the repository adds
WZIP\_S for 4--8\,KiB storage pages.
It starts with the decoded size $n$, which fixes the base schedule and its widest window; a three-byte header then gives the widths the encoder chose for
lengths 3--7, as gaps below the widest. The schedule has entries for lengths 3, 4, 5, 6, 7 and 8+: the power of two of
$n$ selects the base widths of the short lengths, and the widest window is raised to cover the whole input, up to
$2^{27}$ bytes for lengths 8+ and $2^{26}$ for shorter ones. The longest matches thus reach every earlier byte of any
input up to 128\,MiB at no cost in decoder memory, since a one-shot decoder holds the whole output anyway, while small
inputs keep small windows and offset alphabets. For a 16\,MiB input the base widths are 12, 16, 20, 22, 25 and 25 bits,
and the optimal levels, whose parser prices each offset by its code length, widen lengths 3--6 to 18, 22, 24 and 24. Literals are Huffman-coded~\cite{Huffman} in their own stream, in blocks of 32\,KiB;
sequences, in blocks of $2^{14}$, are coded as (i) a \emph{joint symbol} of the offset-cache slot (four recent offsets
or a new one), the literal-run class (0, 1, or 2+ literals) and the length symbol (lengths 3--4095 with extra bits, and
a run token), 1020 values in all, or a 204-value alphabet without the slot, chosen per block; (ii) a literal-run symbol
for runs of two or more; and (iii) for a new offset, a symbol from the Huffman table of the length's group (one group per length below the
first with the widest window, one for the rest; eight finer groups at level 13), plus raw bits. The cache is kept in move-to-front order. The
run token codes a run of the preceding byte with its count in the offset field, bypassing the cache. The joint symbol
captures the dependence of slot, class and length that separate Huffman codes, with their one-bit floor, miss.
A literal block, and each code of a sequence block, either sends its code lengths or reuses the last ones sent, whichever
takes fewer bits (as on a short last block). A block's sequences alternate between two bit streams, so that two decode at once.
An input WZIP cannot shrink by 32 bytes is stored after a zero size field.

\emph{Encoders and decoder.} Level 0 probes one hash table on 6-byte seeds in a 1\,MiB window; levels 1--3 are greedy
and 4--6 lazy, comparing candidates by an estimated gain. Levels 7--13 run the
optimal parser with three indexes: a 3-byte chain for lengths 3--4 searched to $W(4)$, a 5-byte
chain for lengths 5--6 searched to $W(6)$, and a binary tree on 7-byte seeds over the whole window; each chain stops at
a match long enough for the next index or after 16 steps. Each level searches at most its own window: $2^{27}$ bytes
at levels 6 and 13, half per level below, narrower windows staying below it; streams keep the windows of their size,
so decoders need no level. Where matches often reach 1\,MiB back or more, beyond the
L2 cache, the decoder runs the next block of sequences as a pipeline, like Zstandard's long-offset decoder: each step
decodes a sequence, prefetches its source and executes the one decoded 16 steps before (enwik9: 62\% faster).

\emph{Threads.} The indexes depend on the input alone, so each runs ahead of the parser in a thread of its own (the
tree split by hash bucket), with the output of one thread: level 11 compresses Silesia 2.8 times as fast with 6
threads on one core complex. More threads compress 64\,MiB blocks at once, each with the level's window before it as a
dictionary: enwik9 at level 11, 18 times as fast with 105 threads, 0.007\% larger.

\section{Optimal parsing with length-dependent windows}
\label{sec:opt}

Both codecs use the parser of Figure~\ref{fig:opt}, an approximate shortest path over input segments like the
``optimal'' parsers of LZ4HC and Zstandard. Four aspects differ.

\begin{figure}[t]
\centering
\fbox{\begin{minipage}{0.96\textwidth}
\small
\textbf{Parse a segment} from position $p$, with carried literal run $r_0$ and offset cache $R_0$. A node $(c,s)$ is
the cheapest known path from $p$ to $p+c$ in state $s$; it stores price, last step, previous state, pending literal run
and (WZIP) offset cache. A pending run $r$ is in state $\sigma(r)=0$, or (WZIP levels 12--13) in its literal-run
class $\sigma(r)=\min(r,2)$. Initialize $(0,\sigma(r_0))$. For $c=0,1,\ldots$:
\begin{enumerate}
\setlength{\itemsep}{0pt}
\item if $c>0$, for each state $s$ at $c-1$, with pending run $r$, relax $(c,\sigma(r+1))$ by literal $x_{p+c-1}$
and the change in the literal-run code;
\item stop if no path reaches beyond $c$ (if $c>0$), at the segment limit 4096 or at the input end;
\item gather candidates at $p+c$: per-class matches, and (WZIP) each state's cached offsets and a run;
\item for each state $s$, candidate $(L,d)$ and length $\ell\le L$ that $d$ may code, relax $(c+\ell,0)$ with
$\text{price}(c,s)+\text{cost}(\ell,d\mid s)$; (WZIP) its cache is the predecessor's with $d$ moved to the front,
unchanged for a run;
\item on a candidate as long as the level's sufficient length (WZIP: 16--1024), take it from the cheapest state and stop.
\end{enumerate}
Back-trace from the cheapest state at the end, emit, and continue from there.
\end{minipage}}
\caption{\label{fig:opt}The parser of WLZ4's levels 8--12 and WZIP's levels 7--13, a bounded shortest-path search.}
\end{figure}

\emph{Candidates.} A single-window parser can price every length up to the longest match with the nearest candidate
reaching it, the cheapest if offset prices do not decrease with distance. Under a schedule this fails twice. First, a candidate $(L,d)$ codes only the lengths $\ell\le L$ with
$d\le W(\ell)$: a far candidate may not code its own prefixes. WZIP sorts the matches of its three indexes by offset,
keeps each only if it is longer than every nearer one (a heuristic: under Huffman prices a farther offset may cost
less) and admissible at its own length, and prices each length with the nearest such candidate that admits it, skipping lengths no admissible candidate reaches. Second, in a byte format
the offset cost steps with the distance, so WLZ4 keeps one candidate per offset class: the longest match within
128\,B and within 256\,B, within 32\,KiB and within 64\,KiB (each pair from one nearest-first chain walk), and beyond 64\,KiB, pricing each length with the cheapest that codes it. Cached offsets are coded by slot, not bound by the schedule, so
WZIP also tries each at every position with every length, and a run, even of two bytes, as a run token.

\emph{Prices.} WLZ4 prices are exact bytes; each node's pending literal run is charged when the run ends. WZIP prices
symbols in 1/256 bit from smoothed running frequencies, refreshed every 32\,KiB: a match costs its joint symbol, extra
bits, and for a new offset its group's offset symbol and raw bits; the literal-run code is charged incrementally.
The initial counts matter, as the parser's output feeds its prices: a steep prior over lengths makes it split long
matches; WZIP's prior halves every eight.

\emph{States.} The joint symbol codes the literal-run class, so a match's cost depends on how its path arrived:
after a match, one literal, or more. Keeping one path per position commits to one class and misprices the others.
From level 12 WZIP keeps three nodes per position, one per class (a literal moves class $s$ to $\min(s+1,2)$, a match to
0), each with its own offset cache. This remains a bounded dynamic program, not a bit-optimal parse: paths that reach
the same position and class with different caches or pending runs are merged, and segments, candidate limits and
estimated prices restrict it further, as in the other ``optimal'' parsers of practical codecs.

\emph{Passes.} Running prices lag behind changes in the data. At level 13 a first pass (shallow, one state) records
the symbol counts of every 128\,KiB region, and the second pass prices each region from those counts: prices anticipate
the local statistics that the second parse largely reproduces, one step of alternating between parse and code.

\emph{Relation to existing parsers.} LZ4HC prices every length up to one longest match in exact bytes; Zstandard and
LZMA~\cite{LZMA} keep one path per position with its coder state; bit-optimal parsing~\cite{BitOptimal} is exact for fixed costs. New here are per-class
candidates with per-length admissibility, a state per literal-run class and per-region first-pass prices.

\section{Experiments}
\label{sec:exp}

\emph{Setup.} Nodes of two AMD EPYC 9334 (Zen~4, 32 cores each, clock capped at 2.7\,GHz, 32\,MiB of L3 per 8-core
complex), Ubuntu~22.04. Each run is one thread pinned to a core complex of its own, and every configuration ran twice,
on identical nodes, the same day. WLZ4, WZIP, LZ4~1.10.0, Zstandard~1.5.6 (with its assembly Huffman decoder), Brotli~1.1.0 (window
$2^{24}$, its standard maximum) and xz~5.6.2 (no check) were built from source with GCC~14.2 \texttt{-O2}. Each file is
compressed whole, by one library call. Ratio is total input over total output; speeds are total input over total time
after a one-second warm-up, best of 3 compressions for fast codecs (1 otherwise) and of 20 decompressions, over both
runs; every file is verified. LZ4 decodes with its bounds-checked \texttt{LZ4\_decompress\_safe}; WLZ4 and WZIP
decode in their opt-in trusted mode (their default, checked mode decodes Silesia at most 9\% slower). Encoder memory is the growth of the peak resident set during compression,
not counting contexts created beforehand. Corpora: Silesia~\cite{Silesia} (12 files, 5--51\,MB, 211.9\,MB); as small
files, Canterbury and Calgary~\cite{Canterbury,Calgary} (29 files, 3.7\,KB--1\,MB, 6.0\,MB); and as large inputs
enwik8 and enwik9~\cite{enwik}, the first $10^8$ and $10^9$ bytes of an English Wikipedia dump.

\begin{table}[t]
\begin{center}
\caption{\label{tab:lz4class}The LZ4 class: ratio, and compression and decompression speed (MB/s), on Silesia and on Canterbury+Calgary; encoder memory (MiB) on Silesia.}
{\footnotesize
\setlength{\tabcolsep}{4.5pt}
\begin{tabular}{lrrrrrrr}
\toprule
 & \multicolumn{3}{c}{Silesia} & \multicolumn{3}{c}{Canterbury+Calgary} & Memory \\
\cmidrule(lr){2-4}\cmidrule(lr){5-7}
Codec, level & Ratio & Comp. & Dec. & Ratio & Comp. & Dec. & (MiB) \\
\midrule
LZ4 (default) & 2.101 & 735 & 4718 & 2.074 & 627 & 4637 & 0.0 \\
LZ4HC 9 & 2.721 & 38.8 & 4398 & 2.751 & 31.4 & 4550 & 0.7 \\
LZ4HC 12 & 2.743 & 13.4 & 4520 & 2.771 & 4.86 & 4434 & 0.9 \\
\midrule
\textbf{WLZ4 fast} & 2.373 & 328 & 3061 & 2.333 & 290 & 2862 & 0.5 \\
\textbf{WLZ4 lazy} & 2.432 & 307 & 3246 & 2.403 & 275 & 3011 & 0.5 \\
\textbf{WLZ4 2} & 2.614 & 70.6 & 3483 & 2.627 & 68.8 & 3215 & 1.6 \\
\textbf{WLZ4 6} & 2.820 & 40.1 & 3334 & 2.697 & 34.8 & 3448 & 5.2 \\
\textbf{WLZ4 8} & 2.812 & 20.4 & 3333 & 2.773 & 17.5 & 3072 & 7.4 \\
\textbf{WLZ4 10} & 3.009 & 9.21 & 3424 & 2.820 & 10.7 & 3124 & 15 \\
\textbf{WLZ4 12} & 3.186 & 1.71 & 3535 & 2.851 & 2.80 & 3180 & 41 \\
\bottomrule
\end{tabular}}
\end{center}
\end{table}

\begin{figure}[t]
\centering
\includegraphics[width=0.94\textwidth]{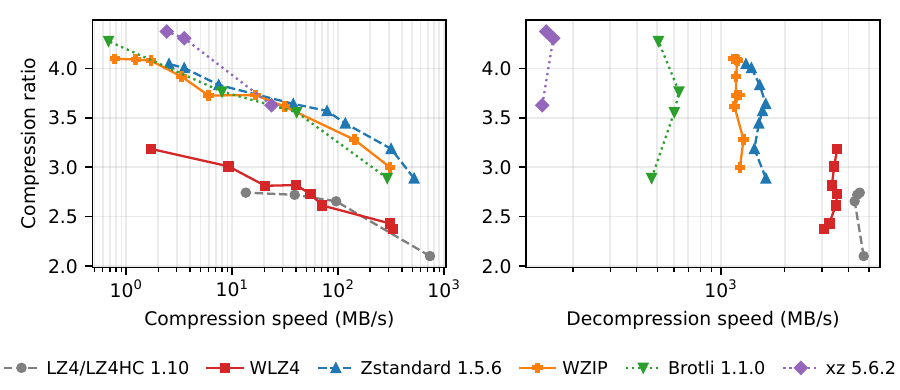}
\caption{\label{fig:tradeoff}Silesia: ratio against compression and decompression speed (log scale).}
\end{figure}

\emph{The LZ4 class} (Table~\ref{tab:lz4class}, Figure~\ref{fig:tradeoff}). On Silesia every WLZ4 level from 6 up
reaches a higher ratio than LZ4HC at its highest level: level 6 reaches 2.820 against 2.743 at 3.0 times LZ4HC-12's
compression speed, and the optimal levels 8, 10 and 12 reach 2.812 (searching 512\,KiB, level 6 4\,MiB), 3.009 and 3.186, up to 16.1\% above it; narrower far
windows cost the lower levels (level 2: 2.614). The fast
and lazy levels gain 13--16\% over LZ4's default mode at 42--45\% of its speed. On the small files
the far window has little to find, and the 32\,KiB two-byte window leaves the hash-chain levels about 2\% below
LZ4HC at similar speed, while levels 10 and 12 still exceed LZ4HC-12 (2.820 and 2.851, against 2.771). The price
is decoding speed: 2.9--3.5\,GB/s, against 4.4--4.7\,GB/s for LZ4. Against Zstandard the
trade-off reverses: level 12 matches Zstandard~3's ratio (3.186, Table~\ref{tab:zstdclass}) and decodes 2.5 times as
fast, but compresses 180 times more slowly.

\begin{table}[t]
\begin{center}
\caption{\label{tab:zstdclass}The Zstandard class: ratio, and compression and decompression speed (MB/s), on Silesia and on Canterbury+Calgary; encoder memory (MiB) on Silesia.}
{\footnotesize
\setlength{\tabcolsep}{4.5pt}
\begin{tabular}{lrrrrrrr}
\toprule
 & \multicolumn{3}{c}{Silesia} & \multicolumn{3}{c}{Canterbury+Calgary} & Memory \\
\cmidrule(lr){2-4}\cmidrule(lr){5-7}
Codec, level & Ratio & Comp. & Dec. & Ratio & Comp. & Dec. & (MiB) \\
\midrule
Zstandard 1 & 2.887 & 521 & 1629 & 3.276 & 420 & 1556 & 0.9 \\
Zstandard 3 & 3.186 & 315 & 1438 & 3.572 & 267 & 1326 & 1.5 \\
Zstandard 19 & 4.005 & 3.52 & 1393 & 4.209 & 4.71 & 1161 & 82 \\
Zstandard 22 & 4.045 & 2.53 & 1312 & 4.210 & 3.55 & 1158 & 642 \\
\midrule
\textbf{WZIP 0} & 2.999 & 308 & 1228 & 3.378 & 236 & 1092 & 1.5 \\
\textbf{WZIP 3} & 3.615 & 31.1 & 1158 & 3.754 & 52.8 & 1105 & 145 \\
\textbf{WZIP 9} & 3.917 & 3.33 & 1178 & 4.147 & 6.02 & 1043 & 174 \\
\textbf{WZIP 11} & 4.078 & 1.71 & 1194 & 4.186 & 3.47 & 1052 & 566 \\
\textbf{WZIP 12} & 4.090 & 1.23 & 1187 & 4.195 & 2.14 & 1052 & 941 \\
\midrule
Brotli 11 & 4.276 & 0.68 & 506 & 4.490 & 0.80 & 470 & 241 \\
xz -6 & 4.305 & 3.52 & 161 & 4.392 & 5.18 & 154 & 94 \\
xz -9e & 4.374 & 2.41 & 149 & 4.396 & 2.49 & 144 & 505 \\
\bottomrule
\end{tabular}}
\end{center}
\end{table}

\emph{The Zstandard class} (Table~\ref{tab:zstdclass}). WZIP's levels 0--9 lie close to Zstandard's trade-off curve on
Silesia but slightly below it (level 9: 3.917 at 3.3\,MB/s; Zstandard 19: 4.005 at 3.5\,MB/s). Its top
levels exceed Zstandard 22's ratio: level 11 reaches 4.078, 0.8\% above Zstandard 22's 4.045, at
68\% of its compression speed, and level 12 reaches 4.090 (+1.1\%); on the small files both fall 0.4--0.6\%
short of it (level 13, not in the table: +0.8\%; Zstandard~1.5.7: within 0.06\%). WZIP decodes 9--29\% slower than
the Zstandard level of nearest ratio (levels 0, 3, 9 and 11--12 against 1, 12, 16 and 22; Figure~\ref{fig:tradeoff}), least
at the top: levels 11 and 12, 1194 and 1187\,MB/s, against Zstandard~22's 1312. The rest lies in the sequences, whose three lookups Zstandard's FSE
states~\cite{Zstd,ANS} make independent and Huffman codes do not (enwik8, decoding alone: 18--22 cycles a sequence,
25--30 with one stream; Zstandard 15--16). WZIP decodes 2.4 times as fast as Brotli 11 and 8.0
times as fast as xz, which compress Silesia further (4.276 and 4.374) at 0.7 and 2.4\,MB/s.
Memory: level 11 (32\,MiB window) 566\,MiB, Zstandard 22 642\,MiB, level 12 941\,MiB.

\begin{table}[!htb]
\begin{center}
\caption{\label{tab:enwik}Large inputs: ratio, and compression and decompression speed (MB/s), on enwik8 (100\,MB) and on enwik9 (1\,GB); encoder memory (MiB) on enwik9.}
{\footnotesize
\setlength{\tabcolsep}{4.5pt}
\begin{tabular}{lrrrrrrr}
\toprule
 & \multicolumn{3}{c}{enwik8} & \multicolumn{3}{c}{enwik9} & Memory \\
\cmidrule(lr){2-4}\cmidrule(lr){5-7}
Codec, level & Ratio & Comp. & Dec. & Ratio & Comp. & Dec. & (MiB) \\
\midrule
LZ4 (default) & 1.746 & 537 & 4333 & 1.964 & 592 & 4426 & 0.0 \\
LZ4HC 12 & 2.386 & 18.3 & 3671 & 2.691 & 20.4 & 3905 & 0.9 \\
\midrule
\textbf{WLZ4 6} & 2.523 & 35.6 & 2980 & 2.859 & 39.5 & 3236 & 5.2 \\
\textbf{WLZ4 10} & 2.766 & 5.80 & 2875 & 3.131 & 6.83 & 3129 & 15 \\
\textbf{WLZ4 12} & 3.046 & 0.88 & 2654 & 3.470 & 1.07 & 2858 & 41 \\
\midrule
Zstandard 1 & 2.459 & 389 & 1425 & 2.803 & 435 & 1544 & 0.8 \\
Zstandard 19 & 3.712 & 2.51 & 1235 & 4.250 & 2.56 & 1366 & 81 \\
Zstandard 22 & 3.957 & 1.53 & 972 & 4.676 & 1.41 & 966 & 649 \\
\midrule
\textbf{WZIP 0} & 2.599 & 221 & 941 & 2.958 & 246 & 1015 & 1.3 \\
\textbf{WZIP 12} & 3.972 & 0.93 & 1032 & 4.645 & 0.90 & 1029 & 1062 \\
\textbf{WZIP 13} & 3.983 & 0.50 & 983 & 4.759 & 0.47 & 950 & 1911 \\
\midrule
Brotli 11 & 3.885 & 0.60 & 534 & 4.477 & 0.62 & 596 & 205 \\
xz -9e & 4.027 & 1.45 & 144 & 4.722 & 1.43 & 165 & 674 \\
\bottomrule
\end{tabular}}
\end{center}
\end{table}

\emph{Large inputs} (Table~\ref{tab:enwik}). Our settings were tuned on Silesia and the small files, and WZIP's
windows for inputs above 64\,MiB on enwik8, the first tenth of enwik9, so enwik9 is not an independent test. WLZ4 12's ratio is 28--29\% above LZ4HC 12's,
but it decodes 27--28\% slower. WZIP 12 (64\,MiB window) exceeds Zstandard 22 by 0.4\% on enwik8 but trails it
by 0.7\% on enwik9, and WZIP 13 (128\,MiB) exceeds it by 0.7\% and 1.8\%; on enwik9 WZIP 13 compresses best
(xz $-9e$: 4.722) and decodes 2\% slower than Zstandard 22, but its encoder uses 1.9\,GiB, Zstandard 22's 0.65.

\emph{Ablation.}\label{sec:ablation} Table~\ref{tab:ablation} compares byte-aligned formats under optimal parsing
(ours, full windows; LZ4HC-12's for LZ4).
On Silesia the far window adds 10--12\% over two windows, and moving it from a match code to the flag adds 2.6--2.8\%
more, since far matches then need no length extension; on the small files the
flagged format's 32\,KiB two-byte window costs 0.2--0.3\%, and flagging lengths 4--5 too adds 0.2\% on both corpora. Assigning offset sizes by length alone lost 1--3\% to the far code on Silesia, 11\% on the small files, where 47--64\% of matches
have length 6+ within 64\,KiB.

WZIP's parts, on Silesia with full windows (\texttt{bench/ablation}; level 11: 4.080): coding the cache slot with the
offset, not in the joint symbol (same parse), lowers the ratio by 0.28\%, and one offset table for all lengths
instead of one per length group (reparsed) by 0.58\%. Giving lengths 3--7 the widest window, with five offset tables
in both runs, changes the ratio by +0.002\% but on mozilla raises peak memory from 0.99 to 1.34\,GiB and
compression time by 11\%: the gain lies in coding offsets per length; the windows keep the indexes small. Levels 12
and 13, whose parsers alone differ from level 11's, add 0.25\% (three states) and 0.17\% more (first pass, eight
groups).

\begin{table}[!htb]
\begin{center}
\caption{\label{tab:ablation}Byte-aligned formats, optimal parsing, full windows: ratio at levels 10 and 12.}
{\footnotesize
\begin{tabular}{lcccc}
\toprule
 & \multicolumn{2}{c}{Silesia} & \multicolumn{2}{c}{Cant.+Calg.} \\
Offset fields & L10 & L12 & L10 & L12 \\
\midrule
LZ4: 2 B, 64\,KiB (LZ4HC-12's parser) & --- & 2.743 & --- & 2.771 \\
1 B 256\,B (len.\ 3--4); 2 B 64\,KiB & 2.749 & 2.769 & 2.772 & 2.789 \\
+ far code: 3 B 16\,MiB, length extension & 3.015 & 3.091 & 2.822 & 2.854 \\
flag instead: 2 B 32\,KiB or 3 B 8\,MiB (len.\ 4+) & 3.095 & 3.179 & 2.815 & 2.846 \\
WLZ4: len.\ 4--5 flagged 1 B 128\,B or 2 B 32\,KiB & 3.101 & 3.186 & 2.820 & 2.851 \\
\bottomrule
\end{tabular}}
\end{center}
\end{table}

\section{Discussion and conclusion}
\label{sec:conclusion}

\emph{Limitations.} WLZ4 decodes 22--32\% slower than LZ4HC~12 on Silesia, and up to 29\% on enwik8 and enwik9, likely
for three reasons: its parses contain more and shorter matches; the flag adds a data-dependent load to each sequence;
and on large inputs a third of its matches copy from 64\,KiB to 8\,MiB back, beyond L2.

\noindent\emph{Conclusion.} Match-length-dependent windows let an LZ77 format give short matches cheap offsets without denying long matches a
large history: as offset fields and a flag in a byte-aligned format, where the windows themselves raise the ratio, and
as offset alphabets coded per length group in an entropy-coded one, where under our parser the gain comes from
the per-length coding and the windows keep the search fast and small. Recording the input length up front costs at
most four bytes and lets the schedule itself adapt to the input. Source code, the benchmark harness and scripts reproducing the
tables are available at \url{https://github.com/icodywu/WZIP/tree/v1.0.0}.

\FloatBarrier
\Section{References}


{\small
\begin{thebibliography}{99}
\setlength{\itemsep}{0pt}
\bibitem{LZ77} J. Ziv and A. Lempel, ``A universal algorithm for sequential data compression,'' \emph{IEEE Trans.
Inf. Theory}, vol. 23, no. 3, pp. 337--343, 1977.
\bibitem{LZ4} Y. Collet et al., ``LZ4: extremely fast compression,'' version 1.10.0, 2024. \url{https://github.com/lz4/lz4}.
\bibitem{DEFLATE} P. Deutsch, ``DEFLATE compressed data format specification version 1.3,'' RFC 1951, 1996.
\bibitem{Zstd} Y. Collet and M. Kucherawy, ``Zstandard compression and the `application/zstd' media type,'' RFC 8878,
2021.
\bibitem{Brotli} J. Alakuijala et al., ``Brotli: A general-purpose data compressor,'' \emph{ACM Trans. Inf. Syst.},
vol. 37, no. 1, art. 4, 2019.
\bibitem{LZMA} I. Pavlov, ``LZMA specification (draft version),'' 2015. \url{https://www.7-zip.org/sdk.html}.
\bibitem{XZ} L. Collin et al., ``XZ Utils,'' version 5.6.2, 2024. \url{https://tukaani.org/xz/}.
\bibitem{ANS} J. Duda, ``Asymmetric numeral systems: entropy coding combining speed of Huffman coding with compression
rate of arithmetic coding,'' arXiv:1311.2540, 2013.
\bibitem{Zstd157} Zstandard v1.5.7 release notes, 2025. \url{https://github.com/facebook/zstd/releases}.
\bibitem{Snappy} Google, ``Snappy format description,'' 2011. \url{https://github.com/google/snappy}.
\bibitem{Lizard} P. Skibi\'nski, ``Lizard v1.x block format,'' 2016. \url{https://github.com/inikep/lizard}.
\bibitem{WuDCC2021} Y. Wu, ``Improved LZ77 compression,'' in \emph{Proc. Data Compression Conf.}, 2021, p. 377.
\bibitem{WuIT} Y. Wu, ``Improved LZ77 compression with match-length-dependent sliding windows,'' submitted to
\emph{IEEE Trans. Inf. Theory}, 2026. arXiv:2610.06530.
\bibitem{Huffman} D. A. Huffman, ``A method for the construction of minimum-redundancy codes,'' \emph{Proc. IRE},
vol. 40, no. 9, pp. 1098--1101, 1952.
\bibitem{BitOptimal} P. Ferragina, I. Nitto and R. Venturini, ``On the bit-complexity of Lempel-Ziv compression,''
\emph{SIAM J. Comput.}, vol. 42, no. 4, pp. 1521--1541, 2013.
\bibitem{Silesia} S. Deorowicz, ``Universal lossless data compression algorithms,'' Ph.D. dissertation, Silesian
University of Technology, 2003.
\bibitem{Canterbury} R. Arnold and T. Bell, ``A corpus for the evaluation of lossless compression algorithms,'' in
\emph{Proc. Data Compression Conf.}, 1997, pp. 201--210.
\bibitem{Calgary} T. C. Bell, J. G. Cleary and I. H. Witten, \emph{Text Compression}. Prentice Hall, 1990.
\bibitem{enwik} M. Mahoney, ``Large text compression benchmark,'' 2006. \url{http://mattmahoney.net/dc/text.html}.
\end{thebibliography}}
\end{document}